\documentstyle[aps,twocolumn,floats,prd,psfig]{revtex}

\newcommand{\ApJL}{Astrophys. J. Lett.}
\newcommand{\ApJ}{Astrophys. J.}
\newcommand{\PRL}{Phys. Rev. Lett.}
\newcommand{\PRD}{Phys. Rev. D}
\newcommand{\MNRAS}{MNRAS}

\newcommand{\aut}[2]{{#2.\ #1}}
\newcommand{\refs}[6]{#2, {\bf #3} {#4} (#5)}

\newcommand{\amp}{and }

\newcommand{\da}{d_A}

\newcommand{\tot}{{\rm t}}

\newcommand{\cmb}{\Theta}
\newcommand{\s}{{\rm s}}
\newcommand{\vecl}{{\bf l}}
\newcommand{\vecla}{{{\bf l}_1}}
\newcommand{\veclb}{{{\bf l}_2}}
\newcommand{\veclc}{{{\bf l}_3}}
\newcommand{\vecld}{{{\bf l}_4}}
\newcommand{\intl}[1]{\int {d^2 {\bf l}_#1 \over (2\pi)^2}}

\newcommand{\bfl}{{\mathbf{l}}}
\newcommand{\dirac}{{\rm D}}
\newcommand{\bp}{{\cal C}}
\newcommand{\shell}{{\rm s}}

\newcommand{\veck}{{\bf k}}

\newcommand{\lens}{{\rm len}}
\newcommand{\pp}{{\phi\phi}}

\newlength{\tskip}
\newlength{\colwidth}

\newcommand{\beq}{\begin{equation}}
\newcommand{\eeq}{\end{equation}}
\newcommand{\beqa}{\begin{eqnarray}}
\newcommand{\eeqa}{\end{eqnarray}}

\newcommand{\deld}{\delta^{\rm D}}
\newcommand{\bn}{\hat{\bf n}}
\newcommand{\bm}{\hat{\bf m}}

\newcommand{\bk}{\hat{\bf k}}
\newcommand{\rad}{r}    

\newcommand{\isw}{{\rm ISW}}

\newcommand{\len}{\phi}

\newcommand{\Ylmn}{Y_{l}^{m}}

\newcommand{\almn}{a_{l m}}

\newcommand{\sz}{{\rm SZ}}

\begin{document}
\twocolumn[\hsize\textwidth\columnwidth\hsize\csname
@twocolumnfalse\endcsname

\title{Weak Lensing of the CMB: Power Spectrum Covariance}
\author{Asantha Cooray\footnote{Sherman Fairchild Senior Research Fellow}}
\address{
Theoretical Astrophysics, California Institute of Technology,
Pasadena, California 91125\\
E-mail: asante@tapir.caltech.edu}

\date{To be submitted to Phys. Rev. D. --- October 2001}

\maketitle


\begin{abstract}
We discuss the non-Gaussian contribution to the power spectrum covariance of 
cosmic microwave background (CMB) anisotropies resulting through weak gravitational lensing angular deflections
and the correlation of deflections with secondary sources of temperature fluctuations
 generated by the large scale structure, such
as the integrated Sachs-Wolfe effect and the Sunyave-Zel'dovich effect.
This additional contribution to the covariance of binned angular power spectrum, 
beyond the well known cosmic variance and any associated instrumental noise, results from a trispectrum, 
or a four point correlation function, in temperature anisotropy data.  
With substantially wide bins in multipole space, 
the resulting non-Gaussian contribution from lensing to the binned power spectrum variance
is insignificant out to multipoles of a few thousand and is not likely to affect the
cosmological parameter estimation with acoustic peaks and the damping tail. 
The non-Gaussian contribution to covariance, however, should be considered when
interpreting binned CMB power spectrum measurements at 
multipoles of a few thousand corresponding to angular scales of few arcminutes
and less.
\end{abstract}
\vskip 0.5truecm

]




\section{Introduction}

The applications of cosmic microwave background (CMB) anisotropy measurements are well
known \cite{Kno95}; its ability to constrain most, or certain combinations of, parameters 
that define the currently favorable cold dark matter cosmologies with a cosmological constant
has driven a wide number of experiments from ground and space. 
The advent of high sensitivity CMB anisotropy experiments with increasing 
capabilities to detect fluctuations over a wide range of scales now suggest the possibility that 
anisotropy power spectrum at small angular scales will soon be measured. 
At angular scales corresponding to few arcminutes and below, 
fluctuations are mostly dominated by secondary effects due to local large scale structure (LSS) 
between us and the recombination. Additionally, important non-linear second order effects such as 
the weak gravitational lensing of CMB leaves important imprints that can in return be used as a probe of cosmology or 
astrophysics related to evolution and growth of structures. 

The increase in sensitivity of current and upcoming CMB experiments also suggest the possibility that non-Gaussian 
signals in the 
CMB temperature fluctuations may be detected and studied in detail. The deviations from Gaussianity in CMB temperature 
fluctuations 
arise through two scenarios: the existence of a primordial non-Gaussianity associated with initial fluctuations and 
the creation 
of non-Gaussian signals through non-linear mode-coupling effects related to secondary contributions. In currently 
favored  cosmologies with adiabatic initial conditions, the primordial non-Gaussianity is non-existent or insignificant 
\cite{KomSpe00}.  This leaves  the secondary contributions, such as the Sunyave-Zel'dovich effect \cite{SunZel80} due to 
inverse-Compton scattering of CMB photons via hot electrons, as the main source of non-Gaussianity.
The existence of non-Gaussian fluctuations in temperature can be directly measured through higher order correlations, 
such as a three-point function or a bispectrum in Fourier space \cite{SpeGol99,CooHu00}.   The detection of non-Gaussianities
at the three-point level can be optimized through the use of
special statistics and matched filters \cite{Coo01a}  
and through certain physical aspects associated with secondary effects, such as frequency dependence \cite{Cooetal00a}.

An additional effect due to non-Gaussianity include a contribution to the four-point correlation function,
or a trispectrum in Fourier space, of CMB temperature fluctuations \cite{Hu00,Zal00}.
The four point correlations are of special interest since they
quantify the sample variance and covariance of two point correlation or 
power spectrum measurements \cite{Scoetal99,CooHu01}.
Thus, to properly understand the statistical measurements of CMB anisotropy fluctuations
at the two point level, a proper understanding of the four point contributions is needed. Similarly, studies of the 
ability of CMB power spectrum measurements to constrain cosmology have been based on a Gaussian 
approximation to the sample variance and the assumption that covariance is negligible. 
If there are significant  non-Gaussian contributions from the four point level that contribute to the power spectrum 
covariance, then it could affect the
conversion of power spectrum measurements to estimates on cosmological parameters. Given the high precision level of
cosmological parameter measurements expected from CMB, a careful consideration must be attached to understanding the 
presence of
non-Gaussian signals at the four point level. Thus, the basic goal of this paper is to understand to what extent 
Gaussian assumption on CMB power spectrum covariance remains to hold when non-Gaussian contributions are included.

As discussed in previous studies (e.g., \cite{SpeGol99}), one of the most important non-linear contribution
to CMB temperature fluctuations is weak gravitational lensing. 
Similar to the weak lensing contribution to CMB anisotropy at the 
three-point level, the contributions to the four point  level results from the non-linear mode-coupling
nature of weak lensing effect and the correlation between
 weak lensing angular deflections and secondary effects that trace the same large scale structure.
The trispectrum due to lensing alone is studied in \cite{Zal00} and the
same trispectrum, under an all-sky formulation, is also considered in \cite{Hu00}.
Here, we focus on the contribution of the trispectrum to the power spectrum covariance which was not 
considered in previous works.  We also include the trispectrum resulting from the correlation between lensing
and secondary effects such as the interagted Sachs-Wolfe (ISW; \cite{SacWol67}) effect and the thermal 
SZ \cite{SunZel80} effect. The latter effect has now been imaged towards massive galaxy clusters where the
temperature of the scattering medium can be as high as 10 keV producing temperature fluctuations of order 1 mK \cite{Caretal96}.

In general, we do not expect secondary contributions
 such as the thermal SZ effect to be an important non-Gaussian contribution to temperature fluctuations, 
since its signal can be easily separated from the thermal CMB spectrum based on multifrequency information \cite{Cooetal00a}.
Still, we consider its correlation with lensing as a source of  covariance as certain experiments, especially at small
angular scales, may not have the adequate frequency coverage for a proper separation. In a previous paper, we discussed 
the non-Gaussian covariance
resulting from SZ alone is discussed in \cite{Coo01b}, where we also considered the effect of full-covariance, compared to
 Gaussian variance assumption, on the estimation of parameters related to the SZ effect. As discussed there, 
due to highly non-Gaussian
behavior of the SZ signal resulting from its dependence on massive halos such as galaxy clusters, the determination of
parameters that define the SZ contribution is affected by the presence of non-Gaussian contribution to the covariance.

In \S \ref{sec:generallensing}, we introduce the basic ingredients for the present calculation.
The CMB anisotropy trispectra  due to weak lensing and correlations between weak lensing
and secondary effects are derived in \S \ref{sec:lensing}.
In \S \ref{sec:covariance}, we calculate the CMB power spectrum covariance due to weak lensing and discuss our results in \S \ref{sec:results}. In \S \ref{sec:summary} we conclude with a summary.

\begin{figure}[t]
\centerline{\psfig{file=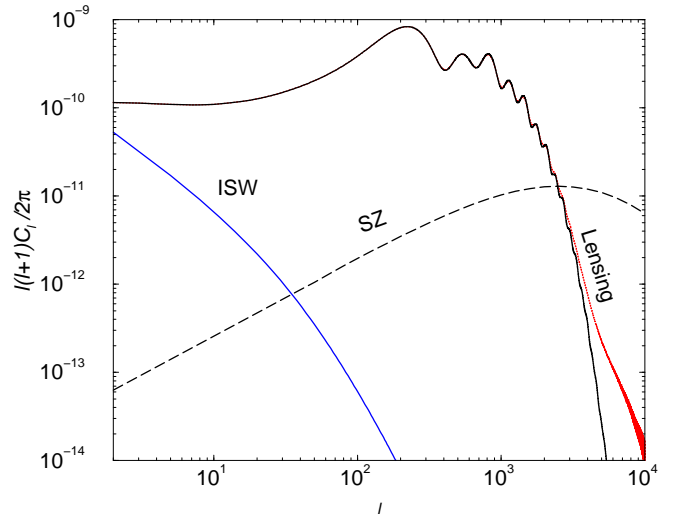,width=3.4in,angle=-90}}
\caption{Power spectrum of the CMB temperature anisotropies in the fiducial $\Lambda$CDM model.
We have also shown the weak lensing, integrated Sachs-Wolfe and the SZ secondary contributions to angular power spectrum.}
\label{fig:cl}
\end{figure}

\section{General Derivation}
\label{sec:generallensing}

Large-scale structure between us and the last scattering surface deflects CMB photons 
propagating towards us. Since lensing effect on CMB is essentially a distribution of photons, from 
large scales to small scales, the resulting effect appears only in the second order \cite{Hu00}.
In weak gravitational lensing, the deflection angle on the sky
given by the angular gradient of the lensing
potential, $\delta(\bn) = \nabla \phi(\bn)$, 
which is itself a projection of the gravitational
potential, $\Phi$ (see e.g. \cite{Kai92}),
\begin{eqnarray}
\phi(\bm)
&=&
- 2 \int_0^{\rad_0} d\rad \frac{\da(\rad_0-\rad)}{\da(\rad)\da(\rad_0)}
                \Phi (\rad,\hat{{\bf m}}\rad ) \,.
\label{eqn:lenspotential}
\end{eqnarray}
The quantities here are the conformal distance or lookback time, from the observer, given by 
\begin{equation}
\rad(z) = \int_0^z {dz' \over H(z')} \,,
\end{equation}
and the analogous angular diameter distance
\begin{equation}
\da = H_0^{-1} \Omega_K^{-1/2} \sinh (H_0 \Omega_K^{1/2} \rad)\,, 
\end{equation}
with the expansion rate for adiabatic CDM cosmological models with a
cosmological constant given by
\begin{equation}
H^2 = H_0^2 \left[ \Omega_m(1+z)^3 + \Omega_K (1+z)^2
              +\Omega_\Lambda \right]\,.
\end{equation}
Here, $H_0$ can be written as the inverse
Hubble distance today $H_0^{-1} = 2997.9h^{-1} $Mpc.
We follow the conventions that
in units of the critical density $3H_0^2/8\pi G$,
the contribution of each component is denoted $\Omega_i$,
$i=c$ for the CDM, $b$ for the baryons, $\Lambda$ for the cosmological
constant. We also define the auxiliary quantities $\Omega_m=\Omega_c+\Omega_b$ and
$\Omega_K=1-\sum_i \Omega_i$, which represent the matter density and
the contribution of spatial curvature to the expansion rate
respectively. Note that as $\Omega_K \rightarrow 0$, $\da \rightarrow \rad$
and we define $\rad(z=\infty)=\rad_0$. Though we present a general derivation of the trispectrum
contribution to the covariance, we show results for the currently favorable $\Lambda$CDM cosmology
with $\Omega_b=0.05$, $\Omega_m=0.35$, $\Omega_\Lambda=0.65$ and $h=0.65$.

The lensing potential in equation~\ref{eqn:lenspotential}  can be related to the well known
convergence generally encountered in conventional lensing studies involving galaxy shear \cite{Kai92}
\begin{eqnarray}
\kappa(\bm) & = &{1 \over 2} \nabla^2 \phi(\bm) \\
            & = &-\int_0^{\rad_0} d\rad \frac{\da(\rad)\da(\rad_0-\rad)}{\da(\rad_0)}
\nabla_{\perp}^2 \Phi (\rad ,\hat{{\bf m}}\rad) \, , \nonumber\\
\nonumber
\end{eqnarray}
where note that the 2D Laplacian operating on $\Phi$ is
a spatial and not an angular Laplacian.
Though the two terms $\kappa$ and $\phi$ contain  differences with respect to
radial and wavenumber weights, these differences cancel
with the Limber approximation \cite{Lim54}. 
The spherical harmonic moments of these two quantities are simply proportional with multiplicative factors in $l$
\begin{eqnarray}
\phi_{l m} &=&
             -{2 \over l(l+1)} \kappa_{l m} =
                 \int d {\bn} \Ylmn{}^*(\bn) \phi(\bn) \nonumber\\
             &=& i^l \int {d^3 {\bf k}\over 2\pi^2} \delta({\bf k})
                \Ylmn{}^* (\bk) I_\ell^{\rm len}(k) \, ,
\label{eqn:GSSZequiv}
\end{eqnarray}
where
\begin{eqnarray}
I_\ell^{\rm len}(k)& =&
                \int_0^{\rad_0} d\rad W^{\rm len}(k,r)
                 j_l(k\rad)  \,,\nonumber\\
W^{\rm len}(k,r)& =&
                -3 \Omega_m \left({H_0 \over k}\right)^2
                F(r) {\da(\rad_0 - \rad) \over
                \da(\rad)\da(\rad_0)}\,.
\label{eqn:lensint}
\end{eqnarray}
Here, we have used the Rayleigh expansion of a plane wave
\begin{equation}
e^{i{\bf k}\cdot \hat{\bf n}\rad}=
4\pi\sum_{lm}i^lj_l(k\rad)Y_l^{m \ast}(\bk)
\Ylmn(\bn)\,,
\label{eqn:Rayleigh}
\end{equation}
and the fact that $\nabla^2 \Ylmn = -l(l+1) \Ylmn$.  

Note that the cosmological Poisson equation relates fluctuations in the density field, given by $\delta$ in 
equation~\ref{eqn:GSSZequiv}, to the fluctuations in the gravitational potential, $\Phi$:
\begin{equation}
\Phi = {3 \over 2} \Omega_m \left({H_0 \over k}\right)^2
        \left( 1 +3{H_0^2\over k^2}\Omega_K \right)^{-2} \frac{G(r)}{a}
        \delta(k,0)\,,
\label{eqn:Poisson}
\end{equation}
with the growth function, $\delta(k,r)=G(r)\delta(k,0)$ \cite{Pee80},
given by 
\begin{equation}
G(r) \propto {H(r) \over H_0} \int_{z(r)}^\infty dz' (1+z') \left( {H_0
\over H(z')} \right)^3\,.
\end{equation}
Note that in the matter dominated epoch $G \propto a=(1+z)^{-1}$.

Expanding the lensing potential to  multiple moments,
\begin{equation}
\phi(\bn) = \sum \phi_{l m} \Ylmn(\bn),
\end{equation}
we can write its correlation function
\begin{eqnarray}
\langle \phi(\hat{\bf n})\phi(\hat{\bf m})\rangle&=&
\sum        \left< \phi_{l m}^* \phi_{l' m'} \right>
           \Ylmn{}^*(\bn) Y_{l'}^{m'} (\bm)\,.
\end{eqnarray}
The power spectrum of lensing potentials is now given through
\begin{equation}
        \left< \phi_{l m}^* \phi_{l m} \right> 	= \delta_{l l'} \delta_{m m'} C_l^\pp
\end{equation}
as
\begin{equation}
C_l^{\phi\phi} = \frac{2}{\pi} \int k^2\, dk P(k)
                I_l^\lens(k) I_l^\lens(k) \,.
\label{eqn:cllens}
\end{equation}

Here, we have introduced the power spectrum of density fluctuations
\begin{equation}
\left< \delta({\bf k})\delta({\bf k')} \right> = (2\pi)^3
        \deld({\bf k}+{\bf k'}) P(k)\,,
\end{equation}
where $\deld$ is the Dirac delta function and
\begin{equation}
\frac{k^3P(k)}{2\pi^2} = \delta_H^2 \left({k \over
H_0} \right)^{n+3}T^2(k) \,,
\end{equation}
in linear perturbation theory.  We use
the fitting formulae of \cite{EisHu99} in evaluating the
transfer function $T(k)$ for CDM models.
Here, $\delta_H$ is the amplitude of present-day density fluctuations
at the Hubble scale; with $n=1$, we adopt the COBE normalization for
$\delta_H$ \cite{BunWhi97} of $4.2 \times 10^{-5}$, consistent with galaxy cluster abundance 
\cite{ViaLid99}, with $\sigma_8=0.86$. 

\begin{figure}[!h]\centerline{\psfig{file=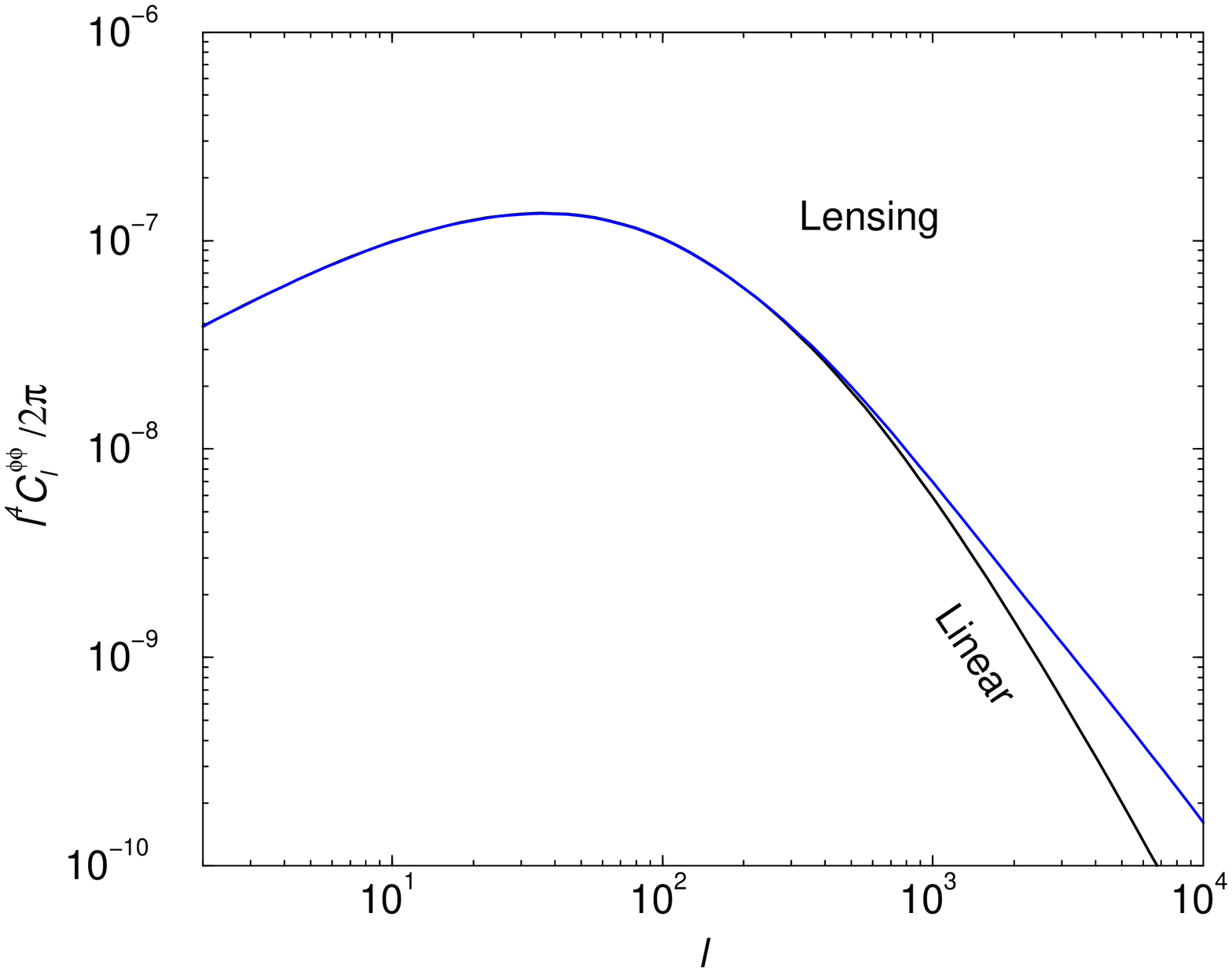,width=3.4in,angle=-90}}
\centerline{\psfig{file=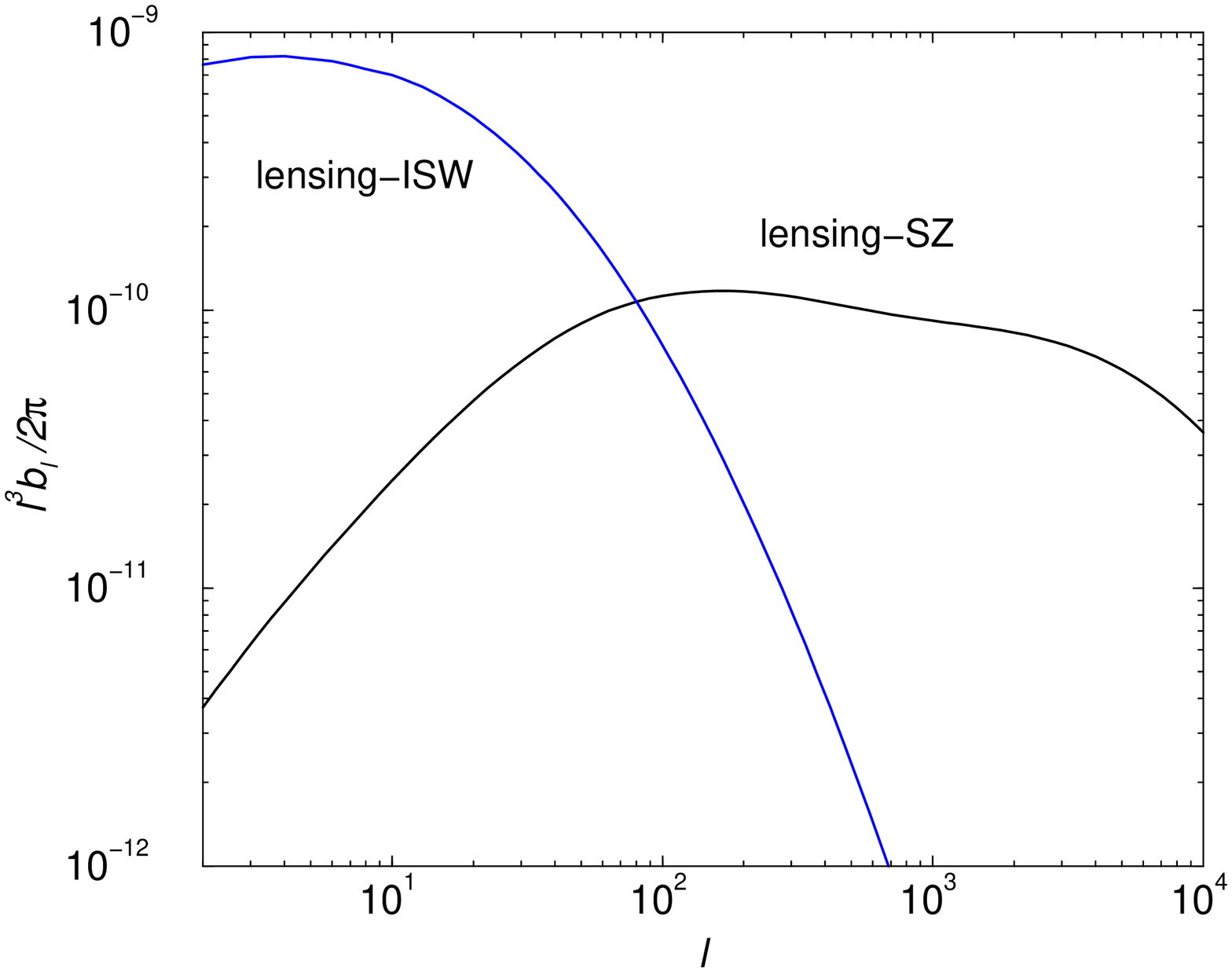,width=3.4in,angle=-90}}
\caption{Power spectra (a) of lensing angular deflections and (b) lensing-isw and lensing-sz cross-correlation. In (a), we show the lensing deflection power spectrum under linear perturbation theory description of the matter fluctuations and using the halo model. As shown,
the lensing deflection power peaks at multipoles of $\sim$ 40 
and act as an effective window function which smooths 
the CMB power spectrum. In (b), the lensing-SZ correlation is calculated with the halo approach while lensing-ISW correlation follows the use of linear theory dark matter power specttrum.}
\label{fig:bl}
\end{figure}

Note that an expression of the type in equation~(\ref{eqn:cllens}) 
can be evaluated efficiently with a version of the Limber approximation \cite{Lim54},
called the weak coupling approximation \cite{HuWhi96},  such that
\begin{eqnarray}
I_l^{\rm X}(k) &\equiv &
                \int_0^{\rad_0}
                 d\rad W^{\rm X}(k,r) j_l(k\rad)  \nonumber\\
          &\approx&
                W^{\rm X}(k,l/k) \int_0^\infty d\rad j_l(k\rad)  \qquad
                (k {W \over \dot W} \gg 1)      \nonumber\\
          &= &
                W^{\rm X}(k,l/k) {\sqrt{\pi} \over 2 k}
                     {\Gamma[(l+1)/2]\over \Gamma[(l+2)/2]}\,.
\label{weakcoupling}
\end{eqnarray}
For $l\gg 1$, the ratio of gamma functions goes to $\sqrt{2/l}$ simplifying further.
Using a change of variables such that $\da=l/k$, we obtain an approximation for the
power spectrum of lensing potentials as
\begin{eqnarray}
C_l^\pp
&=& \frac{2}{\pi} \int k^2\, d k P(k) I_l^\lens(k) I_l^\lens(k) \\
&\approx& \int_0^{\rad_0} \frac{d\rad}{\da^2}\, 
        \left[W^\lens({l \over \da},r)\right]^2
         P\left({l \over\da}\right) \nonumber\, .	
\label{eqn:powerform}
\end{eqnarray}

Since the same large scale structure responsible for deflections in CMB photons produce 
contributions to the anisotropies through other effects, there is a correlation between
the deflection potential and secondary sources of temperature fluctuations. 
As in the power spectrum of delection potential, using statistical isotropy, we write the correlation
as
\begin{eqnarray}
        \left< \phi_{l m}^* \almn^{\rm S} \right> & \equiv & b_l^{\rm S}
         \equiv  {-2\over l(l+1)} C_l^{T \kappa} \,, \nonumber\\
\label{eqn:bl}
         &=& \frac{2}{\pi} \int k^2 dk\, P(k) I_l^{\rm S}(k)
                I_l^{\rm len}(k) \,, \\
         &\approx&
	\int_0^{\rad_0} \frac{d\rad}{\da^2}\,
        W^{\rm S}\left({l \over \da},r\right) W^{\rm len}\left({l \over\da},r\right)
        P\left({ l \over \da}\right) \,. \nonumber
\end{eqnarray}
Here,
\begin{equation}
T^{\rm S}(\bn) = \sum \almn^{\rm S} \Ylmn(\bn),
\end{equation}
and we have used equation~(\ref{eqn:GSSZequiv}) to
relate the power spectrum
$b_l^{\rm S}$ of Refs. \cite{SpeGol99,CooHu00}  and the $\kappa$-secondary
cross power spectrum of Ref. \cite{SelZal99}.  The last line
represents the Limber approximation and we have assumed that
the secondary anisotropies are  related to the first order fluctuations in the density field
projected along the line of sight,
\begin{eqnarray}
a^{\rm S}_{lm} &=& i^l \int \frac{d^3\veck}{2 \pi^2}
\delta(\veck)  I_l^{\rm S}(k) \Ylmn(\hat{\veck}) \, , \nonumber\\
I_l^{\rm S}(k) &=& \int d\rad  W^{\rm S}(k,\rad)j_{l}(k\rad) \, .
\label{eqn:secondaryform}
\end{eqnarray}

In the present paper we consider two secondary effects that correlate with lensing deflections: integrated
Sachs-Wolfe (ISW) effect and the Sunyaev-Zel'dovich (SZ) effect.
Since higher order effects such as the kinetic SZ effect, due to the line of sight motion of
electrons, is second order in density fluctuations, there is no first order cross-correlation with the lensing potentials.
The ISW \cite{SacWol67}  effect results from the
late time decay of gravitational potential fluctuations. The resulting
temperature fluctuations in the CMB can be written as
\begin{equation}
T^\isw(\bn) = -2 \int_0^{\rad_0} d\rad \dot{\Phi}(\rad,\bn \rad) \, .
\end{equation}
The weight function associated with the ISW effect is given by
\begin{eqnarray}
W^\isw(k) = -3\Omega_m \left( {H_0 \over k} \right)^2
         \dot F(r) \, ,
\label{eqn:iswsource}
\end{eqnarray}
which can be used to calculated the correlation between lensing pontential and the ISW effect
through equation~\ref{eqn:bl}.

The SZ effect leads to an effective temperature fluctuation in the
CMB given by the integrated pressure fluctuation along
the line of sight:
\begin{equation}
T^\sz(\bn) = g(x) \int  d\rad  a(\rad)
\frac{k_B
\sigma_T}{m_e c^2} n_e(\rad) T_e(\rad) \,
\end{equation}
where $\sigma_T$ is the Thomson cross-section, $n_e$ is the electron
number density, $\rad$ is the comoving distance, and $g(x)=x{\rm
coth}(x/2) -4$ with $x=h \nu/k_B
T_{\rm CMB}$ is the spectral shape of SZ effect.
At Rayleigh-Jeans
(RJ) part of the CMB, $g(x)=-2$.
For the rest of this paper, we assume observations in the Rayleigh-Jeans
regime of the spectrum; an experiment such as Planck with sensitivity
beyond the peak of the spectrum can separate out SZ contributions
based on the spectral signature, $g(x)$ \cite{Cooetal00a}.

For the correlation between lensing and SZ effect, we follow the halo
model approach of \cite{Coo00}  (see \cite{Sel00} for further details) which
allows a semi-analytical approach to calculate the power spectrum of large scale structure pressure fluctuations.
At RJ part of the frequency spectrum,  the SZ weight function is
\begin{equation}
W^\sz(\rad) = -2 \frac{k_B \sigma_T \bar{n}_e}{a(\rad)^2 m_e c^2}
\end{equation}
where $\bar{n}_e$ is the mean electron density today. With the halo model, we replace the clustering of dark matter
with that of pressure when describing the SZ effect. The cross-correlation between lensing and SZ then involves
the cross-power spectrum between pressure and dark matter.

In figure~\ref{fig:cl}, we show the angular power spectrum of CMB anisotropies \cite{SelZal96}
with secondary contributions through weak lensing, ISW and SZ effects.
The SZ angular power spectrum was calculated using the halo approach of \cite{Coo00}. 
The lensing angular deflection power spectrum and the resulting correlation
power spectra between lensing and ISW, and, lensing and SZ effects
are shown in figure~\ref{fig:bl}.

\section{Lensing Contribution to CMB}
\label{sec:lensing}

In order to derive weak lensing contributions to the CMB trispectrum, we
follow  Hu \cite{Hu00} and Zaldarriaga \cite{Zal00}. We formulate the contribution under a flat sky
approximation; this formulation is adeqaute given that we are mostly interested in
non-Gaussian effects due to lensing at small angular scales
corresponding to multipoles $\gtrsim 1000$. In general, the flat-sky apporach simplifies the derivation and 
computation through the replacement of mode-coupling Wigner symbols through angles.

As discussed in prior papers \cite{SpeGol99,CooHu00,Hu00}, 
weak lensing maps temperature through the angular defelctions resulting along the
photon path by
\begin{eqnarray}
\cmb^\tot(\bn) & = &  \cmb(\bn + \nabla\len) \nonumber\\
        & = &
\cmb(\bn) + \nabla_i \len(\bn) \nabla^i \cmb(\bn) \nonumber\\
&& \quad + {1 \over 2} \nabla_i \len(\bn) \nabla_j \len(\bn)
\nabla^{i}\nabla^{j} \cmb(\bn)
+ \ldots 
\end{eqnarray}
As expected for lensing, note that the remaping conserves the surface brightness distribution of CMB.
Here, $\cmb(\bn)$ is the unlensed primary  component of CMB and $\cmb^\tot(\bn)$ is the total contribution.
It should be understood that in the presence of low redshift contributions to CMB resulting through large scale 
structure, the total contribution includes a secondary contribution which we denote by $\cmb^\s(\bn)$.
Since weak lensing deflection angles also trace the large scale structure at low redshifts, 
secondary effects which are first order in density or potential fluctuations also 
correlate with the lensing deflection angle $\phi$.

Taking  the Fourier transform, as appropriate for a flat-sky, we write
\begin{eqnarray}
\tilde \cmb(\vecla)
&=& \int d \bn\, \tilde \cmb(\bn) e^{-i \vecla \cdot \bn} \nonumber\\
&=& \cmb(\vecla) - \intl{1'} \cmb(\vecla') L(\vecla,\vecla')\,,
\label{eqn:thetal}
\end{eqnarray}
where
\begin{eqnarray}
L(\vecla,\vecla') &=& \len(\vecla-\vecla') \, (\vecla - \vecla') \cdot \vecla'
+{1 \over 2} \intl{1''} \len(\vecla'') \\ &&\quad
\label{eqn:lfactor}
\times \len^*(\vecla'' + \vecla' - \vecla) \, (\vecla'' \cdot \vecla')
                (\vecla'' + \vecla' - \vecla)\cdot
                             \vecla' \,.  \nonumber
\end{eqnarray}

We define the power spectrum and the trispectrum in the flat
sky approximation following the usual way
\begin{eqnarray}
\left< \cmb^\tot(\bfl_1)\cmb^\tot(\bfl_2)\right> &=&
        (2\pi)^2 \delta_\dirac(\bfl_{12}) \tilde C_l^\cmb\,,\nonumber\\
\left< \cmb^\tot(\bfl_1) \ldots
       \cmb^\tot(\bfl_4)\right>_c &=& (2\pi)^2 \delta_\dirac(\bfl_{1234})
        \tilde T^\cmb(\bfl_1,\bfl_2,\bfl_3,\bfl_4)\,. \nonumber \\
\end{eqnarray}

\subsection{Power spectrum}

The power spectrum, according to the present formulation, is discussed in 
\cite{Hu00} and we can write
\begin{eqnarray}
\tilde C_l^\cmb &=& \left[ 1 - \intl{1}
C^{\phi\phi}_{l_1} \left(\vecl_1 \cdot \vecl\right)^2 \right]	\, 
 	                        C_l^\cmb
\label{eqn:ttflat}
\\ && \quad  
        + \intl{1} C_{| \vecl - \vecl_1|}^\cmb C^{\phi\phi}_{l_1}
                [(\vecl - \vecl_1)\cdot \vecl_1]^2  \nonumber\, .
\end{eqnarray}
As written, the second term shows the smoothing behavior of weak lensing through a convolution (eqn.~\ref{eqn:ttflat})
of the CMB power spectrum  (see discussion in \cite{Hu00}). 
With respect to lensing contribution, there are two limiting cases:
when $\vecl - \vecl_1 \approx \vecl$ and cmb power is constant, one can take the CMB power spectrum 
out of the integral in the second term such that
\begin{eqnarray}
\tilde C_l^\cmb &\approx& \left[ 1 - \intl{1}
C^{\phi\phi}_{l_1} \left(\vecl_1 \cdot \vecl\right)^2 \right]\,	
 	                        C_l^\cmb
\\ && \quad  
        + C_l^\cmb \intl{1} C^{\phi\phi}_{l_1}
	                \left(\vecl \cdot \vecl_1\right)^2  \nonumber\, .
\end{eqnarray}
Thus,  there is a net cancellation of terms involving lensing potential power spectrum
and $\tilde C_l^\cmb \approx C_l^\cmb$ producing the well known result that lensing shifts but does not
create power on large scales \cite{Hu00}.   On small scales where there is no or little
intrinsic power in the CMB, the second term behaves such that
\begin{equation}
\tilde C_l^\cmb \approx  {1 \over 2} l^2 C^{\phi\phi}_l \intl{1} l_1^2 C_{l_1}^\cmb   \,.
\end{equation}
Here,  the power generated is effectively the lensing of 
the temperature gradient associated with the damping tail of CMB anisotropy power spectrum. This small angular scale
limit and its uses as a proble of large scale structure density power spectrum and mass distribution of collpased halos 
such as galaxy clusters is considered in \cite{SelZal00}. 

\subsection{Trispectrum}

The calculation of the trispectrum follows similar to the power spectrum. Here, we explicitly show the calculation for
 one term of the trispectrum and add all other terms through permutations.
First we consider the cumulants involving four temperature terms in Fourier space:
\begin{eqnarray}
&& \left< \cmb^\tot(\bfl_1) \ldots
       \cmb^\tot(\bfl_4)\right>_c = \nonumber \\
&& \Big< \left( \cmb(\vecla) - \intl{1'} \cmb(\vecla')
L(\vecla,\vecla')\right) \nonumber \\
&&\quad \times \left(\cmb(\veclb) - \intl{2'} \cmb(\veclb')
L(\veclb,\veclb')\right) \cmb(\veclc)  \cmb(\vecld) \Big>
\nonumber \\
&& \quad =  \Big<
\intl{1'} \cmb(\vecla') L(\vecla,\vecla') 
\intl{2'} \cmb(\veclb') L(\veclb,\veclb') \nonumber \\
&& \quad \times \cmb(\veclc)  \cmb(\vecld) \Big> \, . \nonumber \\
\label{eqn:tri}
\end{eqnarray}
As written, to the lowest order, we find that contributions come from the first order term in $L$
given in equation~(\ref{eqn:lfactor}).
We further simplify to obtain
\begin{eqnarray}
&& \left< \cmb^\tot(\bfl_1) \ldots
       \cmb^\tot(\bfl_4)\right>_c = \nonumber \\
&=&  \Big<
\intl{1'} \cmb(\vecla')
\len(\vecla-\vecla') \, (\vecla - \vecla') \cdot \vecla' \nonumber \\
&& \quad \times
\intl{2'} \cmb(\veclb') 
\len(\veclb-\veclb') \, (\veclb - \veclb') \cdot \veclb'
\cmb(\veclc)  \cmb(\vecld) \Big> \,  \nonumber \\
&=&  C_{l_3}^\cmb C_{l_4}^\cmb \Big<
\len(\vecla+\veclc) \len(\veclb+\vecld) \Big>
\, (\vecla + \veclc) \cdot \veclc \, (\veclb + \vecld) \cdot \vecld \nonumber \\
&& \quad \quad \quad + \, {\rm Perm.}\, , 
\end{eqnarray}
where there is an additional term through a permutation involving the interchange of
$\vecla+\veclc$ with $\vecla+\vecld$. Introducing the power spectrum of lensing potentials, we further simplify to 
obtain the CMB trispectrum due to gravitational lensing:
\begin{eqnarray}    
&&\tilde T^\cmb(\bfl_1,\bfl_2,\bfl_3,\bfl_4) = -C_{l_3}^{\cmb} C_{l_4}^{\cmb} \Big[ 
C^\pp_{|\vecl_1+\vecl_3|} (\vecla +\veclc) \cdot \veclc (\vecla + \veclc) \cdot \vecld  \nonumber \\
&& \quad + C^\pp_{|\vecl_2+\vecl_3|} (\veclb +\veclc) \cdot \veclc (\veclb +\veclc)  \cdot \vecld \Big] 
+ \, {\rm Perm.} \, , 
\label{eqn:trilens}
\end{eqnarray}
where the permutations now contain 5 additional terms with the replacement of
$(l_3,l_4)$ pair by other combination of pairs.


\begin{figure}[!h]
\centerline{\psfig{file=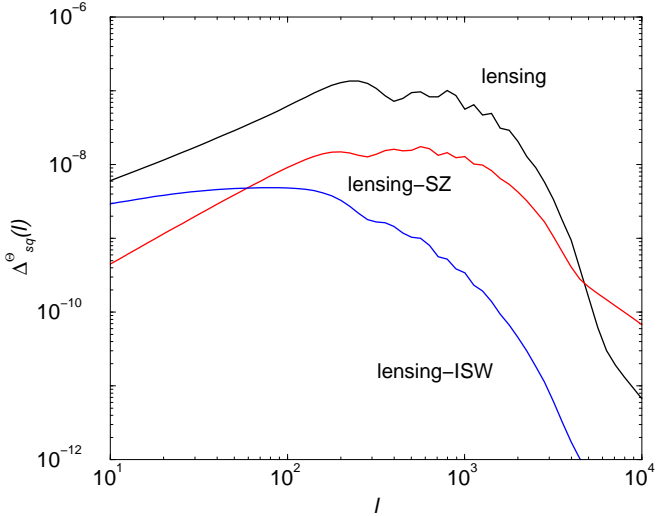,width=3.4in,angle=-90}}
\caption{CMB anisotropy trispectrum resulting from lensing, lensing-ISW and lensing-SZ correlations. The lensing trispectrum generally follows the shape of the CMB power spectrum while lensing-ISW and lensing-SZ trispectra depicts the 
behavior of correlation power between lensing and these secondary effects.}
\label{fig:trispectra}
\end{figure}

\begin{figure}[!h]
\centerline{\psfig{file=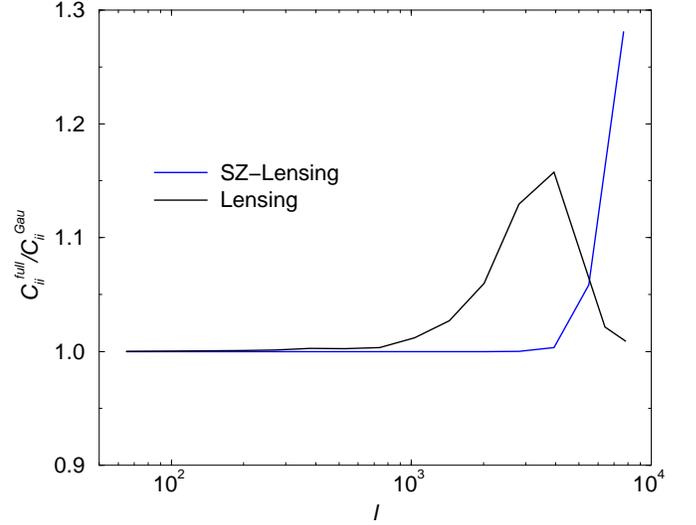,width=3.4in,angle=-90}}
\caption{The ratio of full variance, including non-Gaussian covariance,
to that when non-Gaussian trispectrum contributions are neglected. This
ratio shows the fractional change in the variance, or the errors along the
diagonal of the covariance marix. Out ot multipoles of a few thousand, the
increase due to lensing is less than few percent and at the smallest scale there is no change from the lensing alone trispectrum as its contributions are substatially  small. The lensing-SZ trispectrum only makes noticable contributions at the smallest scale. The resulting changes from the lensing-ISW trispectrum is below $10^{-6}$ and can be ignored.}
\label{fig:variance}
\end{figure}

The non-Gaussian contribution to the trispectrum, through
coupling of lensing deflection angle to secondary effects,
can be calculated with the replacement of $\cmb(\veclc)$ and $\cmb(\vecld)$ in equation~(\ref{eqn:tri}) by
$\cmb^{\rm S}(\veclc)$ and $\cmb^{\rm S}(\vecld)$ 
containing the sources of secondary fluctuations. Thus, 
we can no longer consider cumulants such as $\langle \cmb(\vecla') \cmb^{\rm S}(\veclc)\rangle$ 
as the secondary effects are decoupled from recombination where primary fluctuations are imprinted. However, 
contributions come from the correlation between $\cmb^{\rm S}$ and the lensing deflection $\phi$. 
Here, contributions of equal importance come from both the first and second order terms in $L$ written in
equation~(\ref{eqn:lfactor}).   First, we note
\begin{eqnarray}
&& \left< \cmb(\bfl_1) \ldots
       \cmb(\bfl_4)\right>_c = \nonumber \\
&& \Big< \left( \cmb(\vecla) - \intl{1'} \cmb(\vecla')
L(\vecla,\vecla')\right) \nonumber \\
&\times& \left(\cmb(\veclb) - \intl{2'} \cmb(\veclb')
L(\veclb,\veclb')\right) \cmb^s(\veclc)  \cmb^\s(\vecld) \Big>
\nonumber \\
&& = - C_{l_1}  \left< L(\veclb,-\vecla)\cmb^s(\veclc) \cmb^\s(\vecld) \right> \nonumber \\
&& \quad - C_{l_2} \left< L(\vecla,-\veclb) \cmb^s(\veclc)  \cmb^\s(\vecld) \right> \nonumber \\
&+& \intl{1'} C_{l_1'} \left< L(\veclb,-\vecla')
L(\vecla,\vecla') \cmb^s(\veclc)  \cmb^\s(\vecld) \right> \nonumber \\
\label{eqn:trisec}
\end{eqnarray}
Contributions to the trispectrum from the first two terms come
through the second order term in $L$, with the two $\phi$ terms
coupling to $\cmb^s$. In the last term, contributions come
from the first order term of $L$ similar to the lensing alone contribution to trispectrum.

After some straightforward simplifications, we write the
connected part of the trispectrum involving lensing-secondary coupling as
\begin{eqnarray}    
&&\tilde T^\cmb(\bfl_1,\bfl_2,\bfl_3,\bfl_4) = \nonumber \\
&-& C_{l_3}^{\len\s} C_{l_4}^{\len\s} \Big[  C^\cmb_{l_1} (\veclc \cdot \vecla) (\vecld \cdot \vecla) + 
 C^\cmb_{l_2} (\veclc \cdot \veclb) (\vecld \cdot \veclb) \nonumber \\
&+& \veclc \cdot (\vecla + \veclc)
\vecld \cdot (\veclb +\vecld) C^\cmb_{l_{13}} 
+\vecld \cdot (\vecla + \vecld)
\veclc \cdot (\veclb +\veclc) C^\cmb_{l_{14}}  \Big]  \nonumber \\
&& \quad \quad + \, {\rm Perm.} \, . 
\end{eqnarray}
Note that the first two terms come from the first and second term in equation~(\ref{eqn:trisec}), 
while the last two terms in above are from the third term. As before, through permutations,
 there are five additional terms involving the pairings of $(l_3,l_4)$.

\begin{figure}[!h]
\centerline{\psfig{file=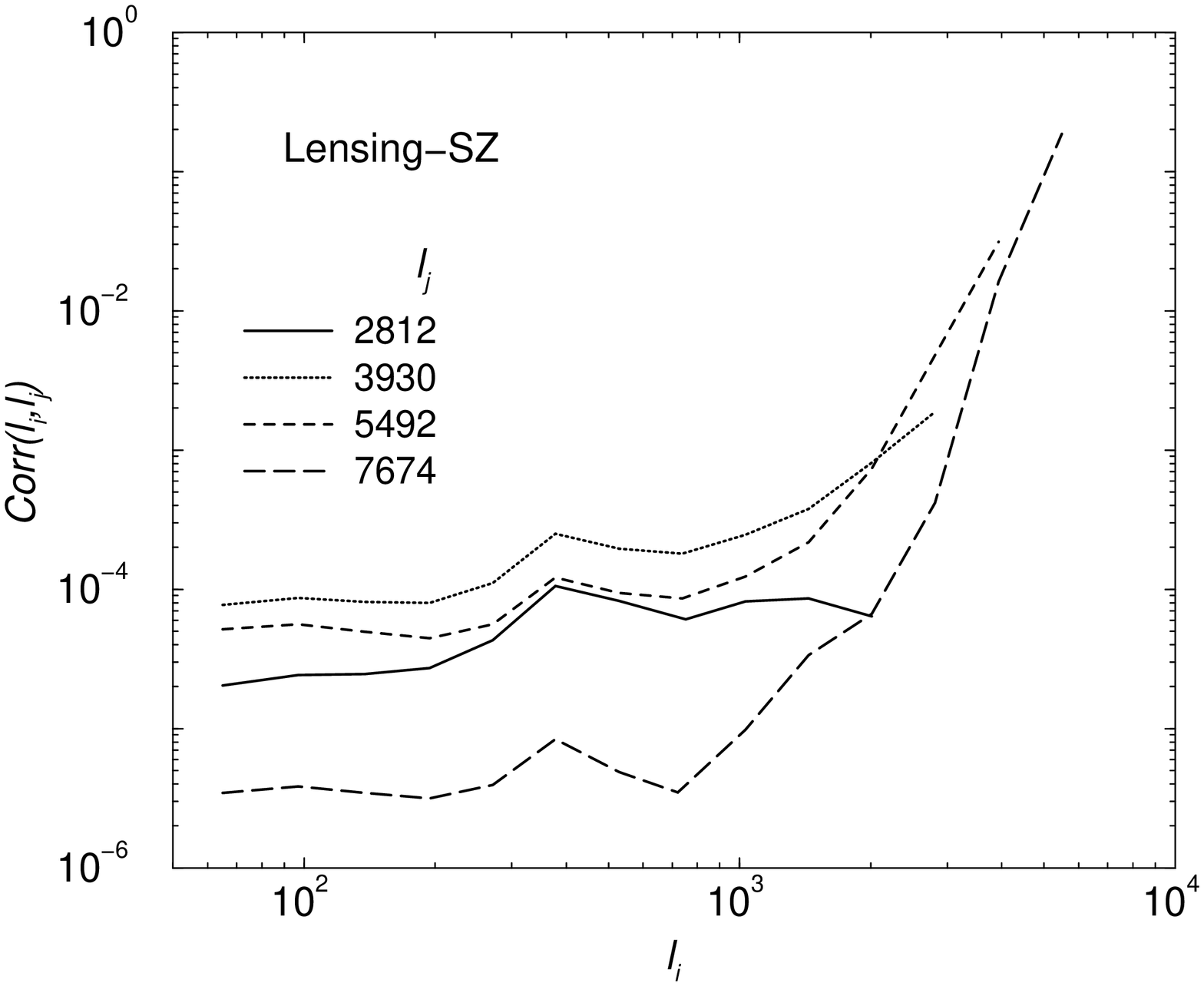,width=3.4in,angle=-90}}
\centerline{\psfig{file=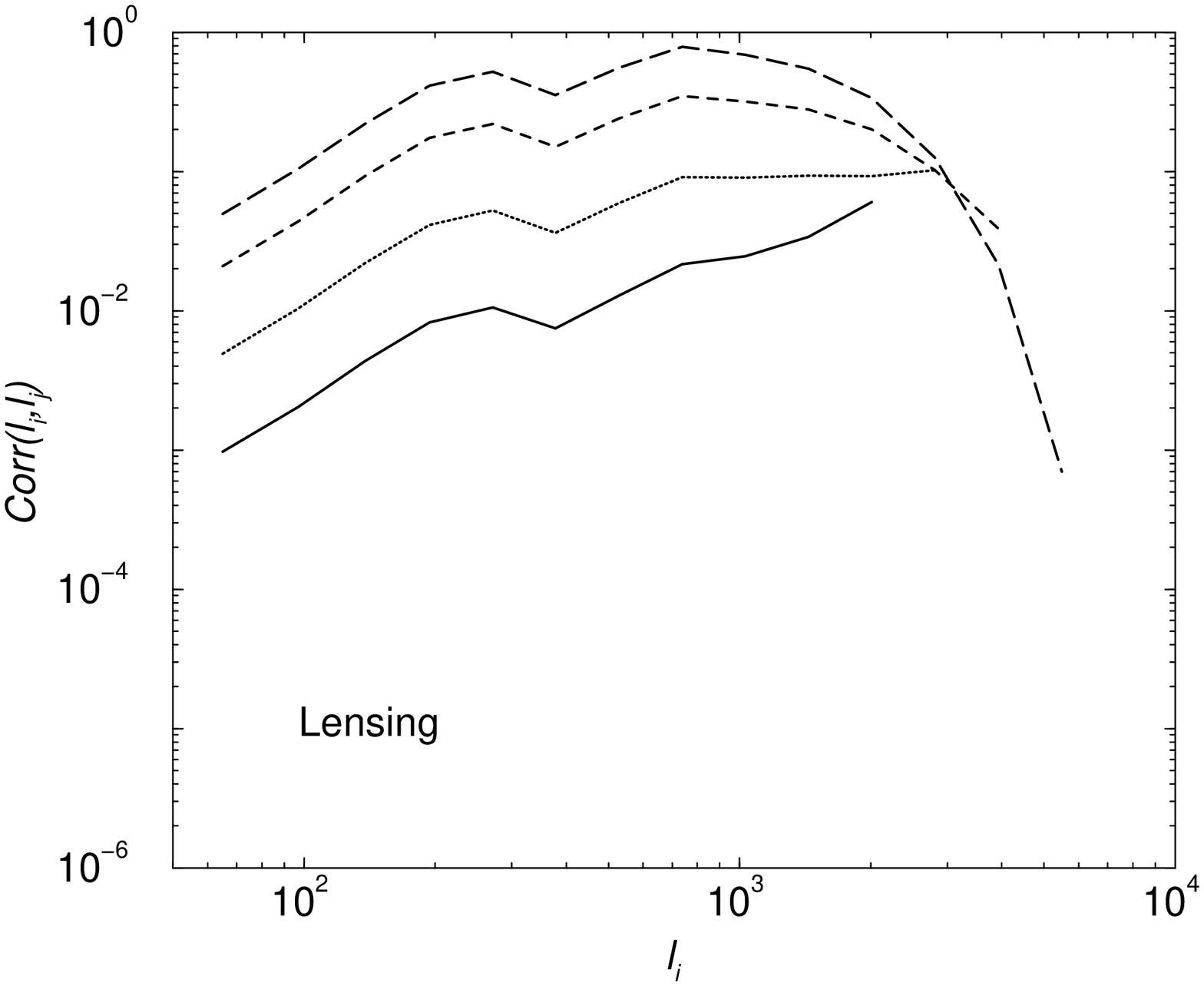,width=3.4in,angle=-90}}
\caption{The correlation coefficients of (a) lensing and (b) lensing-SZ contributions to the covariance of the CMB angular power spectrum. In the case of lensing, the correlations shows the general behavior of lensing effectt on CMB where power is transferred from large angular scales with acoustic peaks to small angular scales, thereby correlating small and large angular scales at the tens of percent level or more. For experiments, such as Planck, that will 
measure the power specrum to multipoles of $\sim$ 2000, the resulting correlations between bins are less than a percent.}
\label{fig:tricorr}
\end{figure}

\section{Power Spectrum Covariance}
\label{sec:covariance}

For the purpose of this calculation, we assume that CMB power spectrum will measure binned logarithmic band
powers at several $l_i$'s in multipole space with bins of thickness $\delta l_i$.
\begin{equation}
\bp_i = \int_{\shell i}
{d^2 l \over{A_{\shell i}}}
\frac{l^2}{2\pi} \cmb(\bf l) \cmb(-\bf l) \, ,
\end{equation}
where $A_\shell(l_i) = \int d^2 l$ is the area of 2D shell in
multipole and can be written as $A_\shell(l_i) = 2 \pi l_i \delta l_i + \pi (\delta l_i)^2$.

We can now write the signal covariance matrix as
\begin{eqnarray}
C_{ij} &=& {1 \over A} \left[ {(2\pi)^2 \over A_{\shell i}} 2 \bp_i^2
+ T^\cmb_{ij}\right]\,,\\
T^\cmb_{ij}&=&
\int {d^2 l_i \over A_{\shell i}}
\int {d^2 l_j \over A_{\shell j}} {l_i^2 l_j^2 \over (2\pi)^2}
T^\cmb(\bfl_i,-\bfl_i,\bfl_j,-\bfl_j)\,,
\label{eqn:variance}
\end{eqnarray}
where $A$ is the area of the survey in steradians.  Again the first
term is the Gaussian contribution to the sample variance  and
includes, in addition to the primary component, contribution
through lensing and secondary effects. The second term is the non-Gaussian contribution.
A realistic survey will also include instrumental noise contributions and we can modify the Gaussian variance to include
the noise through an additional noise contribution to the power spectrum
\begin{equation}
C_l^\tot = C_l^\cmb+N_l\,
\end{equation}
where $N_l$ is the power spectrum of detector and other sources of noise introduced by the experiment.

For the power spectrum covariance, we are interested in
the case when $\veclb = -\vecla$  with $|\vecla|=l_i$ and 
$\vecld = -\veclc$ with $|\veclc|=l_j$. This denotes parallelograms for the trispectrum configuration in multipolar 
or Fourier space. 

In the case of lensing contribution to the trispectrum, with the configuration required for the power
spectrum covariance, we can write
\begin{eqnarray}
&& T^\cmb(\bfl_i,-\bfl_i,\bfl_j,-\bfl_j) = \nonumber \\
&&C_{l_i}^\cmb C_{l_i}^\cmb \left[
C_{|\vecl_i+\vecl_j|}^{\pp} \left[ (\vecl_i+\vecl_j)\cdot \vecl_i\right]^2 +
C_{|\vecl_i-\vecl_j|}^{\pp} \left[ (\vecl_i-\vecl_j)\cdot \vecl_i\right]^2 \right] \nonumber \\
&+&C_{l_j}^\cmb C_{l_j}^\cmb \left[
C_{|\vecl_i+\vecl_j|}^{\pp} \left[ (\vecl_i+\vecl_j)\cdot \vecl_j\right]^2 +
C_{|\vecl_i-\vecl_j|}^{\pp} \left[ (\vecl_i-\vecl_j)\cdot \vecl_j\right]^2 \right] \nonumber \\
&+&2C_{l_i}^\cmb C_{l_j}^\cmb \Big[
C_{|\vecl_i+\vecl_j|}^{\pp}  (\vecl_i+\vecl_j)\cdot \vecl_i(\vecl_i+\vecl_j)\cdot \vecl_j \nonumber \\
&& \quad \quad -
C_{|\vecl_i-\vecl_j|}^{\pp} (\vecl_i-\vecl_j)\cdot \vecl_i (\vecl_i-\vecl_j)\cdot \vecl_j \Big] \, ,
\end{eqnarray}
and includes all terms with no additional permutations.

Similarly, for the lensing-secondary trispectrum we have
\begin{eqnarray}
&& T^\cmb(\bfl_i,-\bfl_i,\bfl_j,-\bfl_j) = \nonumber \\
&&2 \left(\bfl_i \cdot \bfl_j\right)^2 \left[
\left(C_{l_i}^{\len\s}\right)^2 C_{l_j}^\cmb + 
\left(C_{l_j}^{\len\s}\right)^2 C_{l_i}^\cmb \right] \nonumber \\
&-& \left[\left[\bfl_i \cdot (\bfl_i+\bfl_j)\right]^2 
\left(C_{l_i}^{\len\s}\right)^2 
+\left[\bfl_j \cdot (\bfl_i+\bfl_j)\right]^2 
\left(C_{l_j}^{\len\s}\right)^2 \right]C_{|\bfl_i+\bfl_j|}^\cmb \nonumber \\
&-& \left[\left[\bfl_i \cdot (\bfl_i-\bfl_j)\right]^2 
\left(C_{l_i}^{\len\s}\right)^2 
+ \left[\bfl_j \cdot (\bfl_i-\bfl_j)\right]^2 
\left(C_{l_j}^{\len\s}\right)^2\right] C_{|\bfl_i-\bfl_j|}^\cmb \nonumber \\
&+& 2\left[\bfl_i \cdot (\bfl_j-\bfl_i)\right]
\left[\bfl_j \cdot (\bfl_j-\bfl_i)\right]
C_{l_i}^{\len\s}C_{l_j}^{\len\s} C_{|\bfl_j-\bfl_i|}^\cmb \nonumber \\
&-& 2\left[\bfl_i \cdot (\bfl_i+\bfl_j)\right]
\left[\bfl_j \cdot (\bfl_i+\bfl_j)\right]
C_{l_i}^{\len\s}C_{l_j}^{\len\s} C_{|\bfl_i+\bfl_j|}^\cmb \nonumber \\
\end{eqnarray}

\section{Results}
\label{sec:results}

In Fig~\ref{fig:trispectra}, we show the scaled trispectra, where
\begin{equation}
\Delta^\cmb_{\rm sq}(l) = \frac{l^2}{2\pi} T^\cmb(\vecl,-\vecl,\vecl_\perp,-\vecl_\perp)^{1/3} \, .
\end{equation}
and $l_\perp=l$ and $\vecl \cdot \vecl_\perp=0$.
In the case of lensing alone, the trispectrum is proportional to square of the CMB anisotropy power spectrum (see, equation~\ref{eqn:trilens}) and
 the sharp reduction in power at multipoles greater than a few thousand is effectively due to the decrease in
primary anisotropy power at small angular scales.  In the case of lensing-secondary correlation, the trispectrum
is only propertional to one power of the CMB anisotropy power spectrum. Thus, the trispectrum now depicts more of the
behavior of lensing-secondary correlation power spectrum 
shown in figure~\ref{fig:bl}. The sharp decrease in the lensing-ISW trispectrum compared
to that of the lensing-SZ effect is due to differences in the small angular scale power associated with lensing-ISW 
and lensing-SZ correlations.

We can now use this trispectrum to study the
contributions to the covariance.
In Fig.~\ref{fig:variance}, we show the ratio of the diagonal of 
the full covariance to the Gaussian variance with the non-Gaussian term neglected:
\begin{equation}
R \equiv {C_{ii}  \over \bp_i} \, ,
\end{equation}
and for bands $l_i$ given in Table~I. Here, we have used rather wide bins in multipolesuch that bin width is constant in logrithmic intervals in multipole space. This is the same
binning scheme used by \cite{WhiHu99} on $6^\circ \times 6^\circ$ 
fields to investigate weak lensing covariance and later adopted by
\cite{CooHu01}. The two lines show the ratio when 
trispectra due to lensing and 
lensing-SZ correlations are used to calculate the covariance, resepectively.
The square root of the ratio is roughly the fractional change induced in errors along the diagonal resulting from the non-Gaussian
covariance contributions. We do not show the ratio due to lensing-ISW trispectrum as the resulting changes are less than $10^{-6}$ at all
 multipoles of 
interest. As shown in figure~\ref{fig:variance},
the ratio is less than 20\% for weak lensing and peaks at multipoles $\sim 4000$
while the ratio increases to smallest scale with the lensing-SZ
trispectrum.

The correlation between the bands
is given by
\begin{equation}
\hat C_{ij} \equiv \frac{C_{ij}}{\sqrt{C_{ii} C_{jj}}} \, .
\end{equation}
In Table~I we tabulate the correlation coefficients for the
CMB binned power spectrum measurements. 
The upper triangle here is the
correlations under the lensing trispectrum  while the lower triangle shows
the correlations found with the trispectrum due to lensing-SZ correlations.
In the case of lensing contribution to the tripsectrum, correlations
depict the general shape of the CMB power spectrum while in the case of lensing-SZ contribution to the covariance, the correlation coefficients are more consistent with the shape of the lensing-SZ power spectrum.

In Fig.~\ref{fig:tricorr}, we show the correlation coefficients for (a) lensing
and (b)  lensing-SZ contributions to the covariance.
Here we show the behavior of the correlation
coefficient between a fixed $l_j$ (as noted in the figure) 
as a function of $l_i$.  Note that when $l_i=l_j$
the coefficient is 1 by definition; we have not included this point
in the figure due to the apparent discontinuity it creates from the
the dominant Gaussian contribution at $l_i=l_j$.

To better understand how the non-Gaussian contribution scale with our
assumptions,  we consider the ratio of
non-Gaussian variance to the Gaussian variance (see, \cite{Scoetal99,CooHu01})
\begin{equation}
\frac{C_{ii}}{C_{ii}^{\rm G}} = 1 + R \, ,
\end{equation}
with
\begin{equation}
R \equiv \frac{A_{si} T_{ii}^\cmb}{(2
\pi)^2 2 C_i^2} \, .
\label{eqn:rexact}
\end{equation}
In the case of lensing alone contribution to CMB trispectrum, we can simplify the expression for $R$
by noting that
\begin{equation}
T^\cmb(\vecl_i,-\vecl_i,\vecl_j,-\vecl_j)|_{l_i=l_j} \sim 8 C_{l_i}^\cmb C_{l_i}^\cmb C_{\sqrt{2} l_i}^\pp l_i^4 \,
\end{equation}
where for an approximation we have taken $\vecl_i \cdot \vecl_j =0$. 
Replacing the averaging of the product of $(C_{l_i}^\cmb)^2 C_{l_i}^\pp$ with the product of two averages,
we can simplify the ratio of $T_{ii}^\cmb/C_i^2$ to obtain
\begin{equation}
R \sim 4 l_i \delta l_i \left< \frac{l_i^4 C_{l_i}^\pp}{2\pi}\right>_{A_i} \,
\label{eqn:ratio}
\end{equation}
where $< ... >_{A_i}$ represents the averaging of the lensing potential powerspectrum, weighted by a factor of $l_i^4$ to represent
the deflection angles. The equation~(\ref{eqn:ratio}) represents the general behavior of the non-Gaussian contribution to
the lensing trispectrum. The relative contribution from non-Gaussianities
scale with several parameters: (a) increasing the bin size, through
$\delta l_i$ ($\propto A_{si}$), leads to an increase in the
non-Gaussian contribution linearly while (b) the contribution is determined by the shape of the
lensing potential power spectrum. At large multipoles, when $C_{l_i}^\pp$ is a rapidly decreasing function,
there is no significant contribution to the trispectrum and thus to the covariance. This is contrary to the general assumption that
lensing is a small scale phenomena and that small-scale signal in CMB should be strongly affected by the non-linear
nature of weak lensing. In the case of CMB, however, lensing effect is a large scale phenomena, especially given the power spectrum
of lensing deflection angles peak at multipoles of $\sim$ 40 and most of the lensing 
power is concentrated at degree angular scales instead of arcsecond scales.

For upcoming wide-field experiments, especially those involving satellite missions such as MAP and Planck, we do not expect 
non-Gaussianities to limit the interpretation and the cosmological parameter extraction from 
the measured CMB power spectrum. 
For these wide-field experiments, the width of the bin in multipole space will be of order at most few tens;  
for such small bin widths in multipole space at $l \sim$ few hundred  will lead to a significantly reduced 
ratio of non-Gaussian to Gaussian contribution from what we have considered where $\delta l \sim l$.
Also, note that the cosmologically interesting
acoustic peak structure and the damping tail of the CMB anisotropies is limited to multipoles below $\sim$ 1000.
In this range, there is no significant non-Gaussian contribution related to lensing and lensing-secondary correlations.
There are, however, ground-based experiments (e.g., Cosmic Background Imager \cite{Padetal01}) 
for which the non-Gaussianities due to lensing may be important. These small angular scale
experiments, which probe the anisotropy power between multipoles of $\sim$ 1000 and 4000 or so are likely to be limited to
small areas on the sky  and will utilize wide bins in multipole space when estimating the power spectrum in order to
increase the signal-to-noise  associated with its measurement. In such a scenario, it may be necessary to fully account for the
full covariance when interpreting the power spectrum at small angular scales.

\begin{table*}
\begin{flushleft}
\begin{tabular}{cccccccccc}
$\ell_{\rm bin}$
       & 529     & 739   & 1031   & 1440   & 2012 &   2812  &   3930   & 5492  & 7674\\
\hline	
   529 & 1.000 & 0.002 & 0.002 & 0.003 & 0.005 & 0.013 & 0.059  & 0.239& 0.554\\
   739 & (0.000)& 1.000 & 0.010 & 0.007 & 0.009 & 0.021 & 0.091  & 0.348& 0.783\\
  1031 & (0.000)&(0.000)& 1.000 & 0.013 & 0.012 & 0.025 & 0.090  & 0.318& 0.686\\
  1440 & (0.000)&(0.000)&(0.000)& 1.000 & 0.025 & 0.034 & 0.093  & 0.277& 0.547\\
  2012 &(0.000)&(0.000)&(0.000)&(0.000)& 1.000 & 0.060 & 0.092  & 0.200& 0.336\\
  2812 &(0.000)&(0.000)&(0.000)&(0.000)&(0.000)& 1.000 & 0.101  & 0.102& 0.125\\
  3930 &(0.000)&(0.000)&(0.000)&(0.000)&(-0.001)&(-0.002)& 1.000  & 0.039& 0.021\\
  5492 &(0.000)&(0.001)&(0.000)&(0.000)&(-0.001)&(-0.005)&(-0.031)& 1.000& 0.001\\
  7674 &(0.000)&(0.001)&(0.000)&(0.000)&(0.000)&(-0.001)&(-0.016)&(-0.186)& 1.000\\
\hline
\end{tabular}
\caption{ Weak Lensing Convergence Power Spectrum Correlations.
Upper triangle displays the covariance with the lensing trispectrum alone, while the
lower triangle (parenthetical numbers) displays the covariance with the trispectrum due to lensing-SZ correlation.
We do not tabulate the covariance due to trispectrum resulting from lensing-ISW correlation as the correlation coefficients
are of order $10^{-6}$ or below. The data are binned such that bin sizes are constant in logarithmic intervals.}
\end{flushleft}
\end{table*}

\section{Summary}
\label{sec:summary}

The upcoming small angular scale CMB anisotropy experiments are expected to provide first measurements of the power spectrum
related to the damping tail and secondary anisotropies. At such scales, important non-linear effects and secondary contributions can
imprint non-Gaussian signals on the CMB temperature fluctuations. Here, we discussed one aspect related to the presence of 
non-Gaussianities when measuring the CMB anisotropy power spectrum involving a contribution to the covariance resulting from
the higher order four-point correlation function, or a trispectrum in Fourier space.
Here, we discussed the non-Gaussian contribution to the power spectrum covariance of
CMB anisotropies resulting through weak gravitational lensing angular deflections
and the correlation of deflections with the integrated Sachs-Wolfe effect and the Sunyave-Zel'dovich effect.

With substantially wide bins in multipole space,
the resulting non-Gaussian contribution from lensing to the binned power spectrum variance
is insignificant out to multipoles of few thousand containing acoustic peaks and the damping tail, which are of
substantial interest for cosmological parameter estimation purposes. For upcoming satellite based near all-sky experiments, we do not
expect non-Gaussianities to limit the cosmological parameter extraction from  CMB power spectrum measurements and their interpretation.
For small angular scale ground-based  experiments with substantially limited sky coverage, however, the presence
of non-Gaussianities should be accounted when interpreting any measurements at angular scales corresponding to fewe arcminutes
or multipoles $\sim$ 4000. Observational results related to this part of the CMB power spectrum is expected from a wide-variety
of experiments based on interferometric and bolometric techniques. 

\acknowledgments
We acknowledge support from the Sherman Fairchild Foundation and from the DOE grant number
D.E.-FG03-92-ER40701.

\end{document}